\documentclass{optica-article}

\journal{opticajournal} 

\articletype{Research Article}

\begin{document}

\title{Dynamic range extension of an inverse weak value tiltmeter}

\author{Benjamin A. Baldwin\authormark{*} and Garrett K. Josemans}

\address{Naval Surface Warfare Center, Dahlgren Division}

\email{\authormark{*}benjamin.a.baldwin14.civ@us.navy.mil} 


\begin{abstract*} 
We present an investigation into a practical method of extending the dynamic range of an inverse weak value tiltmeter. The investigation is done using a Mach-Zehnder-based, inverse weak value tiltmeter. The tilt information is mapped onto a relative phase-shift between the beam paths. The phase-shift produced by a tilting mirror in one path is negated and tracked by a phase compensator in the other path. Without compensation, the sensitivity of the device drops dramatically for mirror tilts outside of the inverse weak value regime. With compensation, the sensitivity is maintained at a high level over a dynamic range that is extended by approximately four orders of magnitude. 

\end{abstract*}

\section{Introduction}
In 1988, Yakir Aharonov, David Albert, and Lev Vaidman (AAV) defined the weak value and predicted weak value amplification (WVA) \cite{Aharonov:88}; a quantum-mechanical phenomenon which can produce a large magnification of small observable quantities, such as displacement, phase-shift, or tilt. In the years since, the theory has been expanded \cite{Duck:89,Jordan:14,Kofman:12} and confirmed by experiment \cite{Starling:10,Martinez:17} as a viable method for noise suppression and measurement of observables that are undetectably small to other modern methods. For example, an inverse WVA tiltmeter experiment performed at University of Rochester demonstrated nearly shot-nose-limited sensitivity of $200\:\textrm{frad} / \sqrt{\textrm{Hz}}$ \cite{Martinez:17}. To put this in perspective, this corresponds to the breadth of a hair at the distance of the moon \cite{Steinberg:10}.

WVA and inverse WVA \cite{Kofman:12,Starling:10,Martinez:17} offer promising paths to improve existing sensor technologies, but practical applications are just beginning to be explored. The augmentation of common sensors with WVA- and inverse WVA-enhanced readouts is an attractive area of research. Many sensors, specifically those categorized as micro electromechanical systems (MEMS), sacrifice sensitivity for the sake of other considerations such as size, weight, and cost. The sensitivity of MEMS devices is often limited by the readout noise rather than the thermomechanical noise of the device itself. This suggests a possible niche for quantum-hybrid MEMS sensors which would combine the small size and low cost of MEMS devices with the ultra-high sensitivity achievable through WVA and inverse WVA. Such technology would be beneficial to a wide variety of sensors including gyroscopes, gravimeters \cite{Jordan:18}, seismometers, accelerometers \cite{Parks:13}, and more.

Though WVA architectures provide extremely sensitive measurements, they also have limitations. One drawback is their inherently small dynamic range, which limits the applicability of WVA. A proposed solution to this problem is the implementation of a closed-loop feedback mechanism \cite{Haung:10}. In this report, we investigate a particular realization of an inverse WVA tiltmeter with a compensation mechanism that can be implemented in a closed loop. We present theoretical calculations and support them with lab measurements. 

\section{The narrow dynamic range of WVA}

One of the impressive practical benefits of WVA is that it can amplify a signal without amplifying certain types of technical noise. This can enable ultra-sensitive measurements by suppression of the noisefloor \cite{Martinez:17}. In this way, WVA devices can have an advantage over conventional devices. However, by definition, the interaction in a WVA device must be weak, which constrains the upper-limit of the dynamic range.

To understand why this is the case, a brief look at the derivation of the weak value is helpful. Consider a quantum system coupled to a pointer undergoing a weak interaction. For convenience and later reference we choose the quantum system to be a qubit, the pointer state to be a particle’s transverse position along the z-axis, and the weak interaction to be a relative momentum kick between the qubit basis states, $|0\rangle$ and $|1\rangle$. The von Neumann interaction unitary generated by the Hamiltonian for this interaction is,

\begin{equation} \label{eq:unitary}
    U = e^{\frac{i}{\hbar} \frac{k}{2} \hat A \otimes \hat z},
\end{equation}

\noindent
where $k$ is the strength of the relative momentum kick, $\hat z$ is the position operator operating on the pointer state, and $\hat A$ is the $\sigma_3$ Pauli spin operator with eigenvalues ±1 operating on the qubit. The insight of AAV was that, for weak interactions, the following approximations can be made,
\begin{equation} \label{eq:approx}
	\langle f|U|i \rangle 
    \approx \langle f | 1 + \frac{i}{\hbar} \frac{k}{2} \hat A \otimes \hat z | i\rangle
    = \langle f|i \rangle (1 + \frac{i}{\hbar} \frac{k}{2} A_w \hat z) 
    \approx \langle f|i \rangle e^{\frac{i}{\hbar} \frac{k}{2} A_w \hat z},
\end{equation}
\noindent
where $|i\rangle$ and $|f\rangle$ are pre- and post-selected qubit states before and after the weak interaction, and the weak value is a complex number defined as,
\begin{equation} \label{eq:weakvalue}
	A_w = \frac{\langle f|\hat A|i\rangle}{\langle f|i \rangle}.
\end{equation}
It can be seen in Eqs. (\ref{eq:approx}) and (\ref{eq:weakvalue}) that if $|i\rangle$ and $|f\rangle$ are chosen such that $\langle f|i\rangle \ll \langle f|\hat A|i\rangle$, then $A_w$ becomes large and the effect of the interaction unitary is amplified. The limited dynamic range in WVA applications results from the approximations made in Eq. (\ref{eq:approx}). When these approximations fail, WVA theory is not applicable and the effect deteriorates. For an initial pointer state with a Gaussian distribution, the condition that must be met for a weak interaction is,
\begin{equation} \label{eq:condwva}
	k\sigma |A_w| \ll 1,
\end{equation}
\noindent
and for the inverse WVA regime,
\begin{equation} \label{eq:condiwva}
	k\sigma | A_w | \gg 1,
\end{equation}
\noindent
where $\sigma$ is the standard deviation of the Gaussian \cite{Kofman:12}. For specific architectures, Eqs. (\ref{eq:condwva}) and (\ref{eq:condiwva}) can be converted into a rough upper limit on the sensing parameter. This is the limiting factor for the upper-limit of the dynamic range of weak value architectures.  

\begin{figure}[ht!]
\centering\includegraphics[width=8cm]{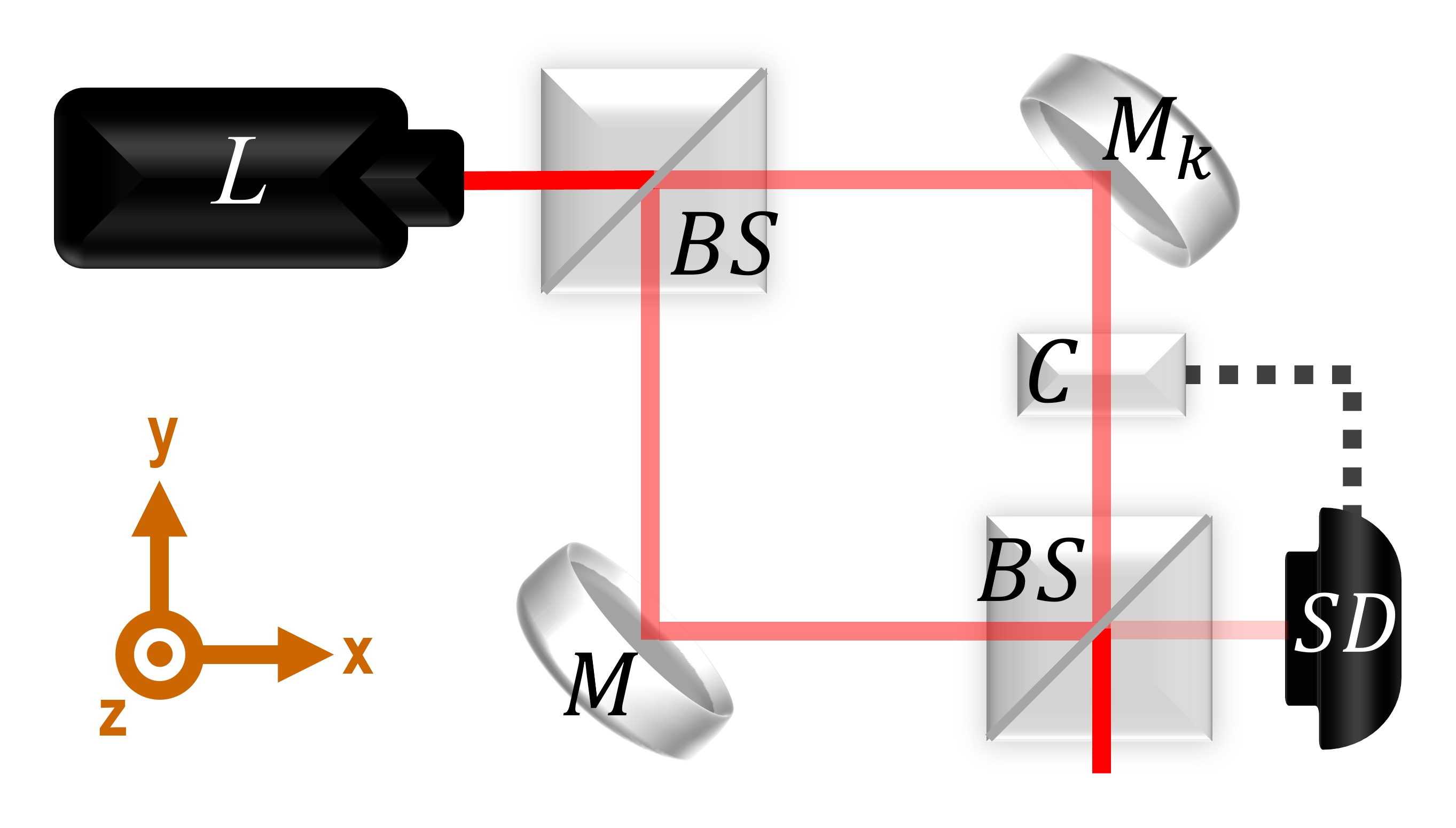}
\caption{MZI-based inverse WVA tiltmeter based on \cite{Martinez:17} with a closed-loop feedback mechanism. A beam is fired from laser $L$ through air into a 50-50 beam splitter $BS$, which separates it into two paths. The upper beam is reflected off mirror $M_k$, which is tilted slightly about a horizontal axis such that the reflected beam's momentum gains a small z-component. The beam then passes through a variable phase compensator which induces a tunable phase shift. The lower beam is reflected off mirror $M$, which is allowed to oscillate about a vertical axis, which is not aligned with the beam's reflection point. This produces an oscillating, relative phase-shift between the beams. The beams are recombined at the second beam splitter and detectors $SD$ monitor the dark port. The dotted line shows information flow from detector outputs to compensator control.}
\label{fig:mzi}
\end{figure}

\section{MZI tiltmeter architecture} \label{sec:architecture}
\begin{figure}[ht!]
\centering\includegraphics[width=10cm]{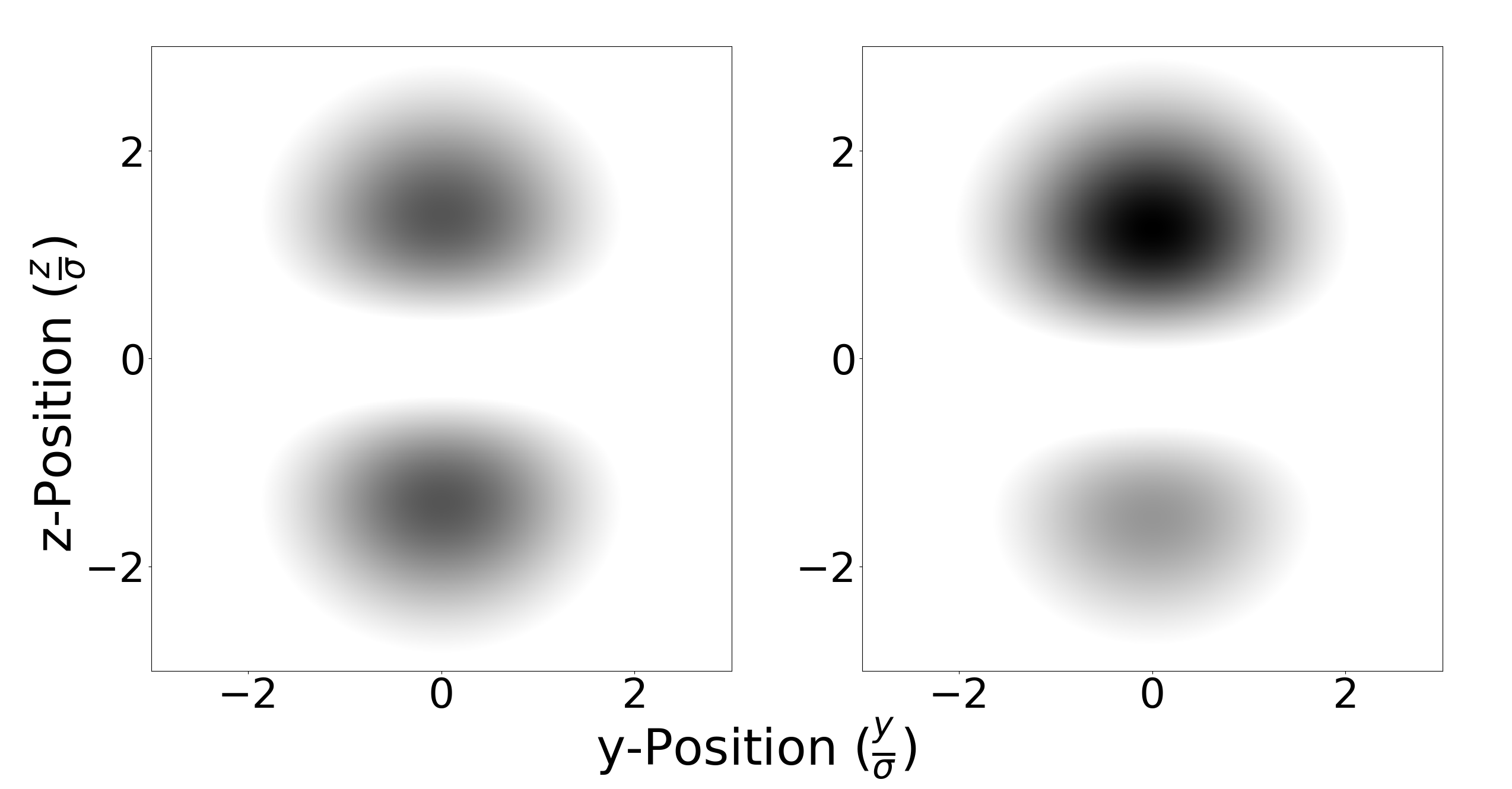}
\caption{Theoretical beam intensity profile at the dark port, from Eqn. (\ref{eq:probdensity}). Darker shading indicates higher intensity, or higher photon probability density. When the upper and lower beams of the MZI are exactly $\pi$ radians out of phase ($\phi = 0$) at recombination, a symmetric dual-mode Gaussian profile is produced (left). As $\phi$ deviates from zero, intensity shifts from one mode to the other (right), resulting in a rapid change in the mean position $\langle z \rangle$ of the beam along the z-axis.}
\label{fig:gauss}
\end{figure}

In 2017, researchers at the University of Rochester introduced an ultrasensitive, optical, inverse WVA tiltmeter \cite{Martinez:17}. The central feature of the Rochester device is a Sagnac interferometer, modified so that the counter-propagating beam-paths are not collinear. In this report, we use a Mach-Zehnder interferometer (MZI) version of the Rochester tiltmeter (Fig. \ref{fig:mzi}). While the overall mathematics that describe the two architectures are nearly identical, the MZI architecture is convenient for inserting optical components into one path without affecting the other. A laser beam enters the MZI at the upper left corner of Fig. \ref{fig:mzi} where it is split between two paths via a 50-50 beam splitter. Along the lower path, the mirror $M_k$ is slightly tilted such that a non-zero angle is introduced between the mirror’s normal vector and the x-y-plane. This reflects the laser beam out of the MZI plane by an angle $\theta_z$, which allows a small amount of light to exit the dark port. The misalignment is very small ($\theta_z \ll 1$) so that the two beams are indistinguishable after recombination. Along the upper path, the mirror $M$ can be rotated by a small angle $\theta$ in the plane of the MZI (about an axis parallel to the z-axis). This introduces an adjustable, relative phase-shift $\phi$ between the counter-propagating beams. The recombined beams create an interference pattern at the dark output-port that takes the form of a symmetric dual-mode Gaussian when $\theta = \phi = 0$ (Fig. \ref{fig:gauss}). When the mirror $M$ is tilted away from its zero-position, $\phi$ becomes non-zero, which causes one mode of the pattern to become brighter, shifting the position expectation value 〈z〉 of the beam spot. The tilt angle  $\theta$ of the mirror $M$ is the sense parameter of the MZI tiltmeter. The tilt information is carried by the phase-shift and mapped onto the mean position $\langle z\rangle$ of the beam spot at the dark port. For small $\phi$, the mean location of the beam is given by $\langle z \rangle \approx 2\phi / k$, where $k = k_o \sin{\theta_z}$ is the momentum kick imparted by the mirror $M_k$ and $k_o$ is the vacuum wavenumber of the laser. Eqs. (\ref{eq:approx}) and (\ref{eq:condiwva}) impose the conditions,
\begin{equation} \label{eq:conds}
    \frac{\phi}{2} \ll k\sigma \ll 1,
\end{equation}
\noindent
on this architecture in order to achieve inverse WVA. As long as these conditions are met, the response of $\langle z \rangle$ to changes in $\theta$ is approximately linear and greatly amplified. These points are made clearer in the quantum treatment in the next section.

\begin{figure}[ht!]
\centering\includegraphics[width=7cm]{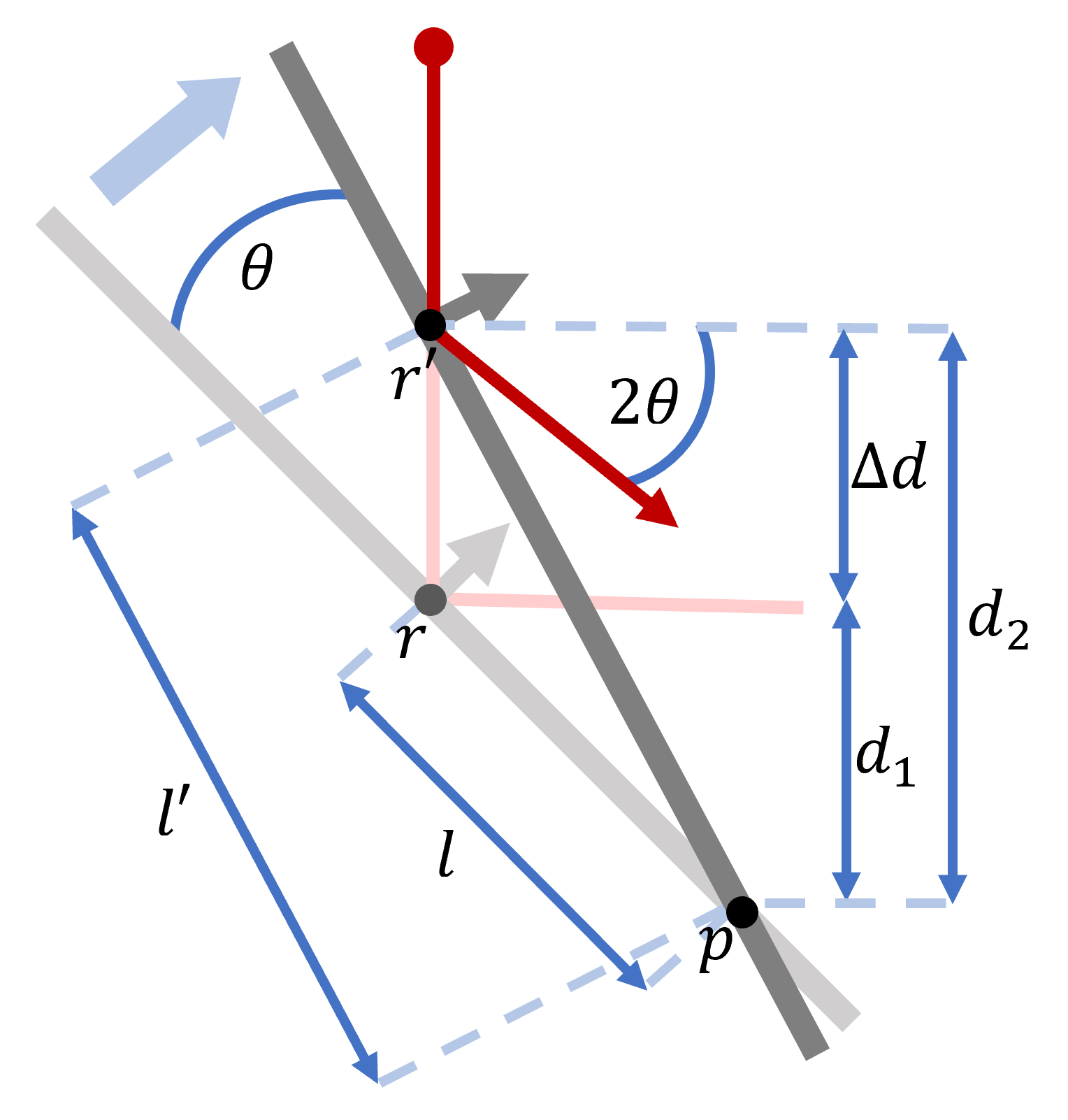}
\caption{Geometry of the change in path-length due to tilt $\theta$ of the tilt of mirror $M$. Note that the tilt angle $\theta$ in this figure is greatly exaggerated for the purpose of illustration. Typical tilts for the reported device are on the order of picoradians and below.}
\label{fig:tilt}
\end{figure}

A relative phase-shift $\phi$ results from the change in the length of the upper MZI path when the mirror $M$ tilts at an angle $\theta$ (Fig. \ref{fig:tilt}). The tilt of the mirror $M$ causes the point of laser reflection to shift from its original position r to its new position $r'$. The total path length change $\Delta d$ is the distance between $r$ and $r'$,
\begin{equation} \label{eq:deltad}
	\Delta d \equiv d_2 - d_1 = \frac{l}{\sqrt2} \bigg[\tan{\bigg( \frac{\pi}{4} + \theta\bigg)}-1\bigg] \approx \sqrt2 l \theta,
\end{equation}
\noindent
where $l$ is the distance between the reflection point $r$ and the pivot point $p$ of the mirror, and the approximation on the right is valid when $\theta \ll 1$. From this, the relationship between phase-shift and mirror tilt is,
\begin{equation} \label{eq:phitotheta}
    \phi = \Delta d k_o = \frac{k_o l}{\sqrt2} \bigg[\tan{\bigg(\frac{\pi}{4}+\theta\bigg)} - 1\bigg] \approx \sqrt2 k_o l\theta,
\end{equation}
\noindent
which is exactly analogous to the Rochester tiltmeter, except here $l$ is the distance from the pivot point of the mirror $M$ to the point of beam reflection rather than the distance between the reflection points of the counter propagating beams of the Sagnac \cite{Martinez:17}.

The change in the angle of incidence between the laser and the surface of the mirror also introduces a misalignment, i.e. a divergence, between the two beams in the x-y-plane. This produces two effects that must be considered. First, the overall path length from $p'$ to the detector is extended by an additional small amount $\Delta D$ such that
\begin{equation} \label{eq:deltaD}
    \cos{(2\theta)} = \frac{D}{D + \Delta D} \rightarrow \Delta D \approx 2D \theta^2
\end{equation}
\noindent
where $D$ is the distance from $r'$ to the detector. Comparing this small length $\Delta D$ to the previously considered path-length change $\Delta d$, it is apparent that, for $\theta \ll 1$, the relative contribution to the phase-shift from this additional term is negligible. Note that $D/l \approx ~ 10\:\textrm{cm}$ for the laboratory setup. Second, the misalignment produced by the mirror $M$ competes with the misalignment produced by the mirror $M_k$. When the inverse WVA conditions in Eq. (\ref{eq:conds}) are satisfied, the interference pattern seen at the dark port is dominated by the misalignment in the z-direction because the tilt $\theta_z$ of the mirror $M_k$ is much larger than the tilt $\theta$ of the mirror $M$.

Though these effects are negligible in the inverse WVA regime, attention is given to them here because it will later be shown that the dynamic-range-extension procedure produces a situation where they become significant.

\section{Quantum description of tiltmeter}
In the MZI tiltmeter, the qubit is the which-path degree of freedom of a photon and the pointer is the z-component of the photon’s transverse position. Upon the photon’s entrance into the MZI, the initial state of the qubit is given by,
\begin{equation} \label{eq:systeminitial}
	|\psi_i\rangle = \frac{1}{\sqrt2} (|L\rangle + |U\rangle),
\end{equation}
\noindent
where $|L\rangle$ and $|U\rangle$ are the basis states representing the possible photon paths through the MZI (“lower” and “upper” according to Fig. \ref{fig:mzi}. The initial distribution of the pointer state is approximately Gaussian,
\begin{equation} \label{eq:meterinitial}
	|\Phi_i\rangle = \int_{-\infty}^{\infty}dz\:|z\rangle\langle z|\Phi_i\rangle
    \:\:\::\:\:\:
    \langle z|\Phi_i\rangle = \frac{1}{\sqrt[4]{2\pi\sigma^2}}e^{-\frac{z^2}{4\sigma^2}},
\end{equation}
\noindent
where $2\sigma$ is the $1/e^2$ beam radius.

On the |L〉 path, a relative momentum kick $k$ is acquired by the photon when reflecting off the mirror $M_k$. Along the $|U\rangle$ path, a relative phase-shift $\phi$ is produced by the mirror $M$’s tilt $\theta$. At the dark port, an additional $\pi$ rad relative phase-shift is produced by the beam splitters. The final qubit state at the dark port is nearly orthogonal to the input state, and is given by
\begin{equation} \label{eq:systemfinal}
	|\psi_f\rangle = \frac{1}{\sqrt2}(|L\rangle - e^{i\phi}|U\rangle).
 \end{equation}
The final state of the pointer at the dark port takes the form,
\begin{equation} \label{meterfinal}
	|\Phi_f\rangle = \langle \psi_f | \hat U | \psi_i\rangle |\Phi_i\rangle = \frac{1}{2} \big(e^{i \frac{k}{2} \hat z} - e^{-i(\frac{k}{2} \hat z + \phi) } \big) |\Phi_i\rangle,
\end{equation}
\noindent
and the un-normalized photon probability density along the z-axis at the dark port is,
\begin{equation} \label{eq:probdensity}
    |\langle z|\Phi_f\rangle|^2 = \frac{1}{2\sqrt{2\pi\sigma^2}} [1 - \cos{(kz+\phi)} ] e^{-\frac{z^2}{2\sigma^2}},
\end{equation}
\noindent
which describes the dual-mode Gaussian intensity profile shown in Fig. \ref{fig:gauss}. The expectation value of the photon’s final position is,
\begin{equation} \label{eq:expectation}
	\langle z\rangle = \frac{\langle \Phi_f |\hat z |\Phi_f \rangle}{\langle \Phi_f|\Phi_f\rangle} = \frac{k\sigma^2 \sin\phi}{e^{\frac{k^2 \sigma^2}{2}}-\cos\phi} \approx \frac{2\phi}{k},
\end{equation}
\noindent
where the approximation on the right is valid when the inverse WVA requirements are met. From Eq. (\ref{eq:expectation}) it can be seen that the position expectation of the laser spot on the detector is proportional to the relative phase-shift $\phi$ and inversely proportional to the weak momentum kick $k$. Thus, the sensitivity of $\langle z \rangle$ to $\phi$ and, from Eq. (\ref{eq:phitotheta}), to $\theta$ is amplified by making $k$ very small.

\section{Phase-shift compensation} \label{sec:compensation}

\begin{figure}[ht!]
\centering\includegraphics[width=7cm]{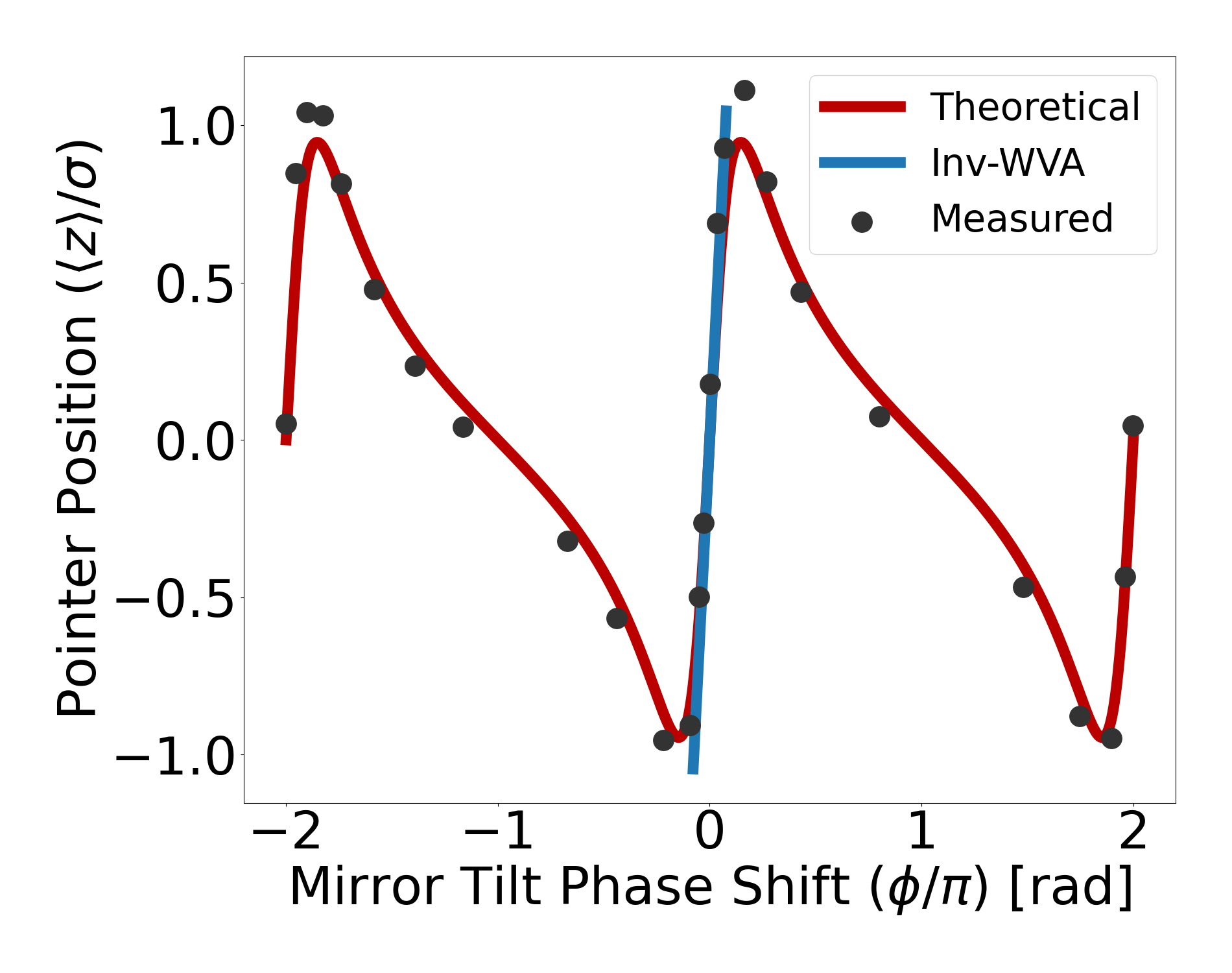}
\caption{Phase-shift dependence comparison between the theoretical pointer position, inverse WVA approximation [both from from Eqn. (\ref{eq:expectation})], and the measured pointer position from the experimental setup.}
\label{fig:iwva}
\end{figure}

The lower-limit of the dynamic range of any device is determined by its noise floor. inverse WVA amplifies the signal but does not amplify the noise, effectively suppressing the noise floor. Smaller values of k result in larger amplification factors and greater noise suppression. However, this also limits how large the sensing parameter can be due to the inverse WVA conditions in Eq. (\ref{eq:conds}). The upper-limit of the dynamic range is determined by these conditions, which, for the MZI tiltmeter, state that the relative phase-shift $\phi$ must be much smaller than the weak interaction strength $k\sigma$. This is made clear by plotting $\langle z \rangle$ and its linear approximation in the inverse WVA regime as functions of $\phi$ from Eq. (\ref{eq:expectation}), as seen in Fig. (\ref{fig:iwva}).

The upper-limit of the dynamic range for this specific application is the point where the inverse WVA approximation fails to match reality within some tolerance (Fig. \ref{fig:iwva}). Though different WVA architectures may depend on different parameters such as linear translation or photon polarization rather than tilt or phase-shift, there will always be an analogous limitation where WVA or inverse WVA approximations fail.

When $|\phi|$ becomes too large, the sensitivity of the pointer to the phase-shift $\textrm d\langle z\rangle / \textrm d \phi$ is no longer approximately linear and the dual-mode Gaussian beam-profile in Fig. \ref{fig:gauss} merges into a single mode. To extend the dynamic range of this architecture, we attempt to keep the effective relative phase-shift at the dark port small by compensating for the phase-shift induced by the mirror $M$. This is done by inserting a variable-phase compensator into the beam-path as seen in Fig. 1. In this configuration, the pointer is sensitive to both the mirror $M$’s tilt $\theta$ and the applied compensation. Because of this, the architecture can be used to characterize both components (the mirror-tilt and the phase-compensation devices) in the ultra-small measurement region (typically below manufacturer specifications) due to the increased sensitivity of the pointer from the inverse WVA effect.

When a phase-shift of magnitude $\phi$ is induced by mirror $M$'s tilt of magnitude $\theta$, the compensator is actuated to produce the same phase-shift in the other beam, maintaining the symmetric dual-mode Gaussian profile on the left of Fig. \ref{fig:gauss}. This keeps the amplified sensitivity $\textrm d \langle z \rangle / \textrm d \phi$ in the inverse WVA regime while allowing the overall tilt $\theta$ to be tracked by the amount of compensation applied.

\section{Experiment} \label{sec:experiment}
The goal of the experiment was to be a proof-of-concept of the dynamic-range-extension mechanism outlined in Section \ref{sec:compensation}. To do so, we aimed to measure the sensitivity of the tiltmeter in- and outside of the inverse WVA regime and compare the results with and without the phase-shift compensation applied.

A 600 nm frequency-stabilized laser (Newport R-32734) produced a linearly-polarized photon-beam, which was fired through air into the input port of the MZI tilt-meter. The mirror $M_k$ was mounted in a piezo-controlled picomotor mount (Newport 8821), which allowed sub-microradian control over $\theta_z$. The mirror $M$ was mounted in a piezo-controlled mirror mount (Thorlabs K2S2P), which was connected to an external function generator (Agilent 33250A) through two -30 dB inline attenuators. The tilt $\theta$ of mirror $M$ was driven by a 1 kHz sinusoidal reference signal from the function generator. The light exiting the dark port of the MZI was monitored by a CCD camera and a quadrant detector (Thorlabs PDQ80A). The top-minus-bottom output of the quadrant detector was fed through a preamplifier (SRS SR560) set to an amplification factor of 1000 with a band-pass filter centered at 1 kHz with a 6 dB/oct rolloff. The preamp output was connected to an oscilloscope (Tektronix MDO3014) which was connected to a computer. 

The compensation mechanism was implemented by inserting a Soleil-Babinet Compensator (Thorlabs SBC-VIS) into the lower beam-path. The SBC was oriented such that the beam polarization was parallel to the optic axis of one of the internal plates and perpendicular to the other. In this orientation, the polarization of the beam was unaffected by the SBC, but the internal, variable-width crystal could be adjusted to fine-tune the phase-shift of the lower beam-path. The response of the MZI tiltmeter output to applied SBC actuation was found to be linear and consistent over many phase-cycles.

A baseline (BL) configuration was established by blocking the lower beam of the MZI, eliminating the inverse WVA effect. In this configuration, the transverse photon distribution at the output was approximately Gaussian. The normalized, right-minus-left output of the quadrant detector is given by,
\begin{equation} \label{eq:quaddetout}
	\mathcal O_{BL} = \frac{\int_0^\infty \textrm dx\: \psi(x) - \int_{-\infty}^0 \textrm dx\: \psi(x)}{\int_{-\infty}^\infty \textrm dx\: \psi(x)}
    = \frac{ \textrm{erf}(D \sin{(2\theta)})}{\sqrt2\sigma} 
    \approx \frac{2\sqrt2 D \theta}{\sqrt\pi\sigma},
\end{equation}
\noindent
where $\psi(x) = e^{-\frac{(x-\langle x\rangle)^2}{2\sigma^2}}, \langle x\rangle = D\sin{(2\theta)}$ was the offset of the Gaussian peak due to the mirror $M$’s tilt $\theta$, and $D$ was the path-length from the tilting mirror to the detector. The linear approximation on the right of Eq. (\ref{eq:quaddetout}) is good over the range of $\theta$ values relevant to this report. The linear response $R=\mathcal O_BL/V_{in}$ of the quadrant detector output amplitude $\mathcal O_BL$ to the amplitude of the mirror $M$’s drive signal $V_{in}$ was measured by driving the mount at 1 kHz for various input amplitudes. This was necessary due to the fact that $R$ is frequency dependent because of physical effects in the mirror-mount’s piezo actuator. The output waveforms from the quadrant detector were recorded and transformed to the frequency domain via FFT, from which the output amplitudes were found. Using the measured value of $R$ and Eqs. (\ref{eq:phitotheta}) and (\ref{eq:quaddetout}), the phase-shift as a function of drive-signal amplitude is given by,
\begin{equation} \label{eq:phitov}
    \phi = k_o \frac{l}{\sqrt2} \bigg[\tan{\bigg(\frac{\pi}{4} + \sin^{-1}{\bigg(\frac{\sigma \textrm{erf}^{-1}(RV_{in})} {D \sqrt 2} \bigg)} \bigg)} - 1 \bigg] \approx \frac{\sqrt\pi k_o \sigma lR}{2D V_{in}}.
\end{equation}
The same procedure was repeated with the beam block removed to determine the amplified response of the device in the inverse WVA regime for different $k\sigma$ values.

\begin{figure}[ht!]
\centering\includegraphics[width=7cm]{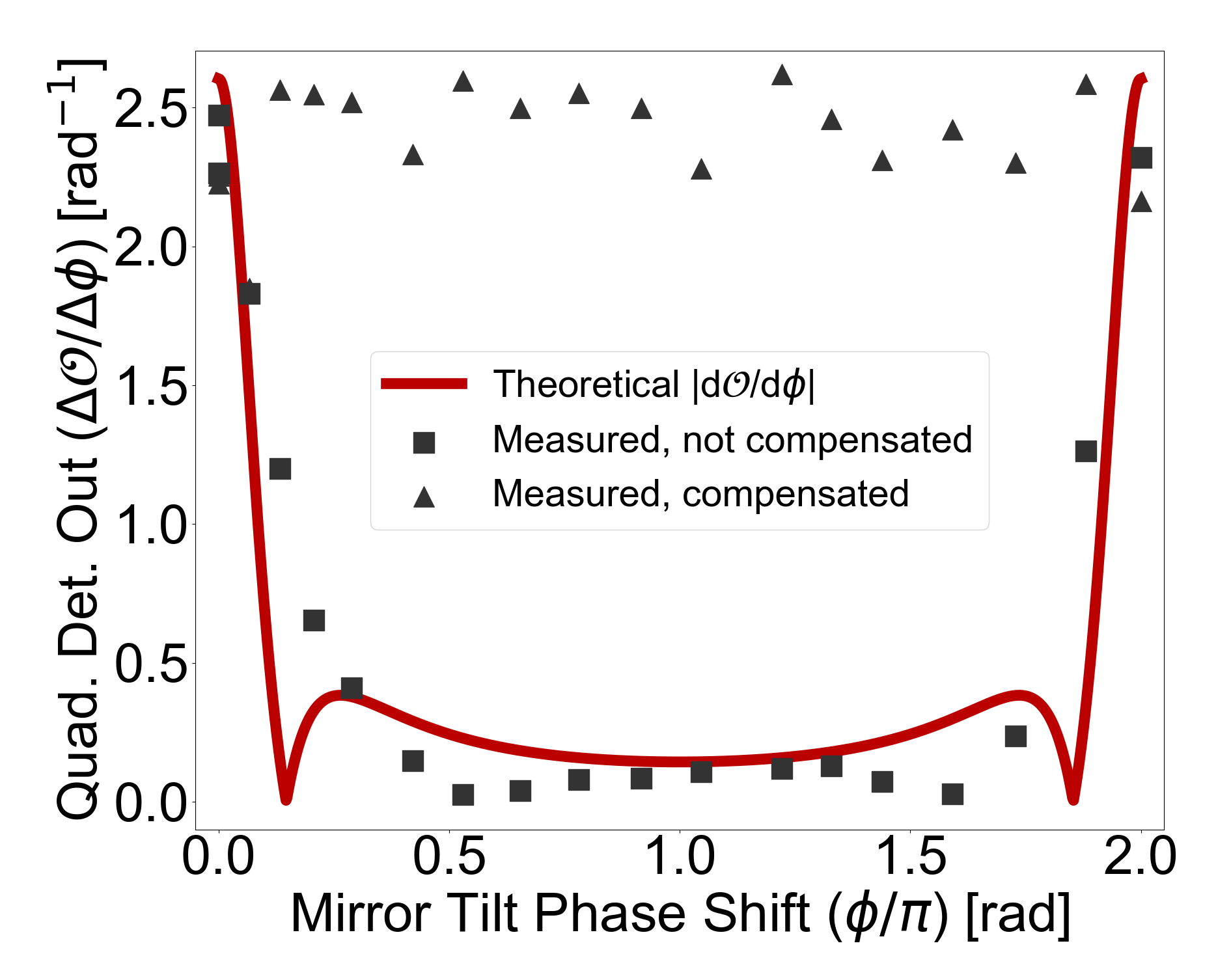}
\caption{Pointer sensitivity of the MZI tiltmeter with $k\sigma = .47$; a comparison of compensated and non-compensated trials to theory. The non-compensated trials show significant decrease in sensitivity outside the inverse WVA regime, which defines the dynamic range of the tiltmeter. The compensated trials maintain high sensitivity over a much larger range because the phase compensation keeps the pointer in the inverse WVA regime.}
\label{fig:result}
\end{figure}

The normalized, top-minus-bottom output $\mathcal O$ of the quadrant detector takes the form,
\begin{align} \label{eq:normalquaddetout}
	\mathcal O = \frac{\int_0^\infty \textrm dz\: |\langle z|\Psi_f\rangle|^2 - \int_{-\infty}^0 \textrm dz\: |\langle z|\Psi_f \rangle |^2}{\int_{-\infty}^{\infty} \textrm dz\:|\langle z|\Psi_f\rangle|^2} 
    = \frac{\textrm{erf}^{-1}\bigg(\sqrt{\frac{k^2 \sigma^2}{2}}\bigg) \sin\phi}{e^{\frac{k^2 \sigma^2}{2}} - \cos \phi } 
    &= \frac{\textrm{erf}^{-1}\bigg(\sqrt{\frac{k^2 \sigma^2}{2}}\bigg)}{k\sigma^2} \langle z \rangle \nonumber
    \\
    &\approx \sqrt{\frac{2}{\pi\sigma^2}} \frac{2\phi}{k},
\end{align}
\noindent
where the approximation on the right is valid in the inverse WVA regime. It can be seen in Fig. 4 that outside of this regime the sensitivity of the output signal to phase-shift $\textrm d \mathcal O / \textrm d\phi$ is greatly diminished. From Eq. (\ref{eq:normalquaddetout}), this sensitivity is given by,
\begin{equation}
    \frac{\textrm d \mathcal O}{\textrm d \phi} = \textrm{erf}^{-1}    \bigg(\sqrt{\frac{k^2 \sigma^2}{2}}\bigg) \bigg[\frac{\cos\phi}    {e^{\frac{k^2 \sigma^2}{2}} - \cos\phi} - \bigg(\frac{\sin\phi}    {e^{\frac{k^2 \sigma^2}{2}} - \cos\phi} \bigg)^2 \bigg].
\end{equation}
\noindent
The value of $\textrm d \mathcal O /\textrm d\phi \approx \Delta \mathcal O / \Delta \phi$ was measured by driving the piezo-controlled mirror $M$ with a sinusoidal reference signal which induced an oscillating phase-shift of amplitude $\Delta\phi$ which was determined from the amplitude of the drive signal $V_{in}$ and Eq. (\ref{eq:phitov}). The value of $\Delta \mathcal O$ was the amplitude of the induced quadrant-detector-output signal, determined from the FFT of the waveform. This was repeated at various values of $\phi$ (over a full $2\pi$ rad phase-cycle) by applying DC offset voltages to the mirror $M$'s mount which tilted the mirror $M$ at various angles. The same procedure was then repeated using the SBC to cancel-out the tilt-induced phase-shift from the mirror $M$, forcing the relative phase-shift $\phi$ to remain in the inverse WVA regime for each measurement (Fig. \ref{fig:result}). The loop between the detector and the SBC was closed by the experimenter observing the output and manually adjusting the SBC to compensate. The overall tilt $\theta$ of the mirror $M$ was tracked by the amount of SBC actuation needed to cancel the phase-shift.

\section{Results}

Without applying phase-compensation, measurements of $\Delta \mathcal O / \Delta \phi$ in the inverse WVA MZI tiltmeter roughly follow the theoretical curve for $|\textrm d \mathcal O / \textrm d \phi|$. The upper limit of the dynamic range of this device is loosely set by the relation $\phi/2 \ll k\sigma$. For $k\sigma = .47$, the measured values of $\Delta \mathcal O / \Delta \phi$ dropped by $\sim$26\% over the first 0.2 rad phase-shift interval and continued to decline rapidly. Over the interval $\sim1 < \phi < \sim5$ rad, the measured value of $\Delta \mathcal O / \Delta \phi$ never exceeded ~5\% of the maximum, measured at $\phi = 0$.

With phase-compensation applied, $\Delta \mathcal O / \Delta \phi$ remained roughly constant near the maximum value over a full $2\pi$ rad phase-cycle. Variations were observed over this interval due to the extreme sensitivity of the device and the susceptibility of mirror $M$'s mount to random drift as well as air currents and external vibrations. This resulted in difficulty holding the mirror at a fixed angle $\theta$ long enough to obtain clean measurements. Nevertheless, the trend of the measured data clearly demonstrates that high sensitivity was maintained over the full phase-cycle. Furthermore, the trend was quantitatively observed to continue over several phase-cycles in each direction (greater than $\pm10\pi$ rad). 

After shifting through several phase-cycles, the integrity of the beam profile observed at the detector began to deteriorate. More specifically, the dual-mode Gaussian profile shown in Fig. \ref{fig:gauss} began to rotate such that its axes of symmetry were no longer parallel to the x- and z-axes. This was due to the buildup of the effects pointed out at the end of Section \ref{sec:architecture}, particularly in Eq. (\ref{eq:deltaD}), where the accumulation of tilt at mirror $M$ produced a momentum kick in the x-y plane that began to compete with the one produced at the mirror $M_k$. Though not quantified in this report, this effect places a boundary on the upper-limit of the extended dynamic range for this particular architecture.

\section{Conclusion}
We built an MZI version of the Sagnac inverse WVA tiltmeter introduced by researchers at the University of Rochester in \cite{Martinez:17} and demonstrated that the device can produce great amplification of small tilts and measured a noisefloor of $\sim$1 prad/$\sqrt{\textrm{Hz}}$ at a tilt-oscillation frequency of 1 kHz.

We have demonstrated in theory and supported with experimental observation that the dynamic range of the inverse WVA MZI tiltmeter can be extended by the implementation of the closed-loop phase compensation method described above. Experimentally, the upper-limit of the dynamic range was extended from a few mrad of phase-shift to several tens of radians. The largest source of experimental uncertainty was from environmental effects on the piezo-controlled mirror mount used to produce the reference tilt-signal. Implementation of closed-loop phase-compensation using automatic, electronic feedback and control mechanisms should provide finer control and measurement than the human-in-the-loop feedback mechanism implemented in this study.

Though a mirror and an SBC were used to produce the phase-shifts in this report, it should be noted that this device is agnostic to the source of the phase-shifts. Thus, like all interferometers, the device is sensitive to any process that produces a change in the effective path-length of one of the arms. The amplification of the pointer-shift at the detector from the inverse WVA effect is the additional advantage that allows this device to achieve a higher level of sensitivity to the underlying cause of the phase-shifts.

The limited dynamic range imposed by WVA is a significant hurdle to realizing the benefits of WVA in practical devices. The results presented in this report are a step toward that end.

\begin{backmatter}

\bmsection{Funding}
These efforts were funded under Office of Naval Research (ONR) award N0001423WX00070.

\bmsection{Acknowledgments}
Grateful acknowledgement is given to ONR whose sponsorship and continuing guidance of the In-House Laboratory Independent Research program has made this work possible. The authors also thank the Technology Office of Naval Surface Warfare Center, Dahlgren Division for their dedication and advocacy of this research.

\bmsection{Disclosures}
The authors declare no conflicts of interest.

\bmsection{Data availability}
Data underlying the results presented in this paper are not publicly available at this time but may be obtained from the authors upon reasonable request.

\end{backmatter}


\bibliography{bib}






\end{document}